\documentstyle[11pt]{article}
\begin{document}
{\newcommand{\bU}{\mbox{\boldmath $\rm U$}}
{\newcommand{\bM}{\mbox{\boldmath $\rm M$}}
{\newcommand{\bV}{\mbox{\boldmath $\rm V$}}
{\newcommand{\eD}{\end{document}}
{\newcommand{\re}{{\rm e}}
{\newcommand{\ri}{{\rm i}}

{\Large{\bf  May We Expect }}\hfill TUD-IKTP/99-01\\
{\Large{\bf CP- and T-Violating Effects}}\\ 
{\Large{\bf in Neutrino
Oscillations?}}\hfill 1 Feb 1999\\[4mm]

K.\ R.\ Schubert\\ Institut f\"{u}r Kern- und Teilchenphysik\\
Technische Universit\"{a}t Dresden\\
 Schubert@physik.tu-dresden.de\\[4mm]

{\small {{\bf Abstract:} Neutrino oscillations with three families of
leptons are described with the help of a unitary neutrino mixing
matrix {\bf U} in analogy to the Standard Model quark mixing with
the CKM matrix. If {\bf U} contains a nontrivial phase and if all
three neutrino mass eigenstates have a different mass, there will
be CP- and T-violating asymmetries at distances $L$ which are
large compared to $E/D^2$, where $D^2$ is the largest difference
of neutrino mass squares. We give explicit expressions for these
$L$-dependent asymmetries in a
frame which is able to describe the present solar and atmospheric
oscillation observations.}}\\[4mm]

 The recent strong evidence for neutrino oscillations in
atmospheric neutrino production \cite{SUPERK} has amplified the
interest in the basic properties of these oscillations. The
observed strength of the neutral weak interaction of atmospheric
neutrinos, in the $\pi^0/\rm e$ ratio \cite{SUPERK-2},
 does not require the assumption of
sterile neutrinos but allows a complete description in an extended
Standard Model of elementary particles.\\

Without changing anything in the principles of the model, the
extension can easily be made by adding three SU(2) singlets of
right-handed  neutrinos
--- $\nu_{eR}$, $\nu_{\mu R}$, and $\nu_{\tau R}$ --- to
the set of elementary fermions. This leads to a fourth mass matrix
$\bM_{\nu}$ in addition to $\bM_\ell$, $\bM_Q$, and $\bM_q$ for charged
leptons $\ell$, up- and down-type quarks $Q$ and $q$. 
The diagonalization of
$\bM_\ell$ leads to the Yukawa term $m_e {\rm{\overline e}e}+m_\mu
{\overline\mu}\mu+m_\tau{\overline\tau}\tau$ in the Lagrangian,
and $\bM_\nu$ cannot be diagonalized at the same time.
The SU(2) partners of e, $\mu$, and $\tau$ have to be 
linear superpositions of mass eigenstates $\nu_1$,
$\nu_2$, and $\nu_3$. In very close analogy to the CKM matrix of
quark mixing \cite{CABIBBO,KM}, the right-handed neutrino singlet
fields require a unitary transformation $\bU$ between the two triplets
in family space, 
$$\nu^\prime=\left( \begin{array}{c}\nu_e\\
\nu_\mu\\ \nu_\tau\\\end{array}\right)\ {\rm and}\ \ \nu=\left(
\begin{array}{c}\nu_1\\ \nu_2\\ \nu_3\\\end{array}\right).$$
For experimental convenience, $\bU$ should be defined for the
neutrinos and not for the charged leptons, i.\ e.
 $$\nu^\prime=\bU\ \nu,\ \ \bU =\left( \begin{array}{ccc}U_{e1}&
U_{e2}&U_{e3}\\U_{\mu 1}&U_{\mu 2}&U_{\mu 3}\\
U_{\tau 1}&U_{\tau 2}&U_{\tau 3}\\ \end{array}\right),
\ \  \nu=\bU^\dagger\ \nu^\prime,
 \ \ \overline\nu^\prime=\bU^*\ \overline\nu,
 \ \ \overline\nu=\bU^{\rm T}\ \overline\nu^\prime ,$$
in contrast to the quarks where it is completely arbitrary if $\bV$
describes mixing of $q$ and $q^\prime$ with mass-diagonal $Q$ or
mixing between $Q$ and $Q^\prime$ with diagonal $q$ fields. Since the
first publication of Kobayashi and Maskawa \cite{KM} we are used
to the choice $q^\prime=\bV q$, but the opposite convention is
preferred for leptons since only their charged components can be
directly detected.\\

The extended Standard Model has 25 free parameters. In addition to
the conventional 18, these are $$m(\nu_1),\ m(\nu_2),\ m(\nu_3),
\ \vartheta_{e2},\ \vartheta_{\mu 3},\ \vartheta_{e3},
\ {\rm and}\ \Delta,$$ where we have parametrized $\bU$ as
$$\bU=\left(
\begin{array}{ccc}1 & 0 & 0\\0 & c_{\mu 3} & s_{\mu 3}
\\0 & -s_{\mu 3} & c_{\mu 3}
\\ \end{array}\right) \left(
\begin{array}{ccc}c_{e3} & 0 & s_{e3}e^{-i\Delta}
\\0 & 1 & 0\\-s_{e3}{e^{i\Delta}} & 0 & c_{e3}\\ \end{array}\right)\left(
\begin{array}{ccc}c_{e2} & s_{e2} & 0\\-s_{e2} & c_{e2} & 0\\0 & 0 & 1 
\end{array}\right)$$ with $0\le\vartheta_{ij}\le\pi/2$,
$c_{ij}=\cos\vartheta_{ij}$,
$s_{ij}=\sin\vartheta_{ij}$, and $0\le\Delta <2\pi$ in
complete analogy to the parametrisation of $\bV$ by the Particle Data
Group \cite{PDG}. If all three masses $m(\nu_i)$ are different
from each other, if all three mixing angles are different from 0
and from $\pi/2$, and if $\Delta$ is different from 0 and from
$\pi$, it is impossible \cite{JARLSKOG} to get a real matrix $\bU$ by
rotating lepton field phases. The extended Standard Model then
gets the possibility of showing T-violating and CP-violating
effects in the lepton sector, the strengths of which are
proportional to the phase-rotation invariant quantity \cite{JARLSKOG}
$$J^{(\nu)}\ =\ s_{\mu 3}\ s_{e3}\ s_{e2}\ c_{\mu 3}\ c_{e3}^2
\ c_{e2}\ \sin\Delta\ .$$
For any $i,f,k,l$ (without summing), this quantity is
given by \cite{JARLSKOG}
$$J^{(\nu)}\ =\ {\rm Im}(U_{ik}U_{fl}U_{il}^*U_{fk}^*)\ \cdot
\ \sum_{\alpha,\beta}\ \epsilon_{if\alpha}\epsilon_{kl\beta}\ .$$ 
 T-and CP-violating effects require
these phase-rotation invariant
combinations of four mixing matrix elements in the pertinent
amplitudes. We investigate their possible presence in the
following.\\

At its time of production by a W-boson, a neutrino is given by
$$\nu_i^{\prime}=\sum_{j=1}^3U_{ij}\ \nu_j$$ with $i=e,\ \mu$, or
$\tau$. After a distance $L$, this neutrino disappears by creation
of a charged lepton $f=e,\ \mu$, or $\tau$ . The probability for
this process is $$P(i\to
f)\ =\ P_{fi}\ =\ |\sum^3_{j=1}U_{ij}U_{fj}^*\ \re^{-\ri m_j^2L/2E}|^2.$$ At
small distances, $P_{fi}=\delta_{fi}$ because of $\bU\bU^\dagger=1$. We now
order the mass eigenstates by increasing mass and abbreviate
$$m_2^2-m_1^2=d^2,\ \ m_3^2-m_1^2=D^2.$$ 
Present observations by
Super-Kamiokande \cite{SUPERK} and by the various solar neutrino
disappearance experiments \cite{SOLAR} indicate
$$d^2\approx5\cdot10^{-6}{\rm eV}^2\ll D^2\approx3\cdot 10^{-3}
{\rm eV}^2\ ,$$
where the value for $d^2$ follows from the best fit with the small angle
solution including the MSW effect
\cite{MSW}.\\

 For a given neutrino energy $E$, the first distances 
with observable
oscillations are of the order $L=o(L_2)$ with $$ L_2\ =\ E/D^2.$$
At these distances, we can approximate $d^2\approx 0$  and get
$$P_{fi}=|U_{i1}U_{f1}^*+U_{i2}U_{f2}^*+U_{i3}U_{f3}^*\ \re^{-\ri D^2L/
2E}|^2.$$
Because of the unitarity of $\bU$, this is
$$P_{ii}=|(1-|U_{i3}|^2)+|U_{i3}|^2 \re^{-\ri D^2L/2E}|^2$$
$$=1-4|U_{i3}|^2(1-|U_{i3}|^2)\sin^2(D^2L/4E),$$ and for $f\neq i$
 $$P_{fi}=|U_{i3}|^2|U_{f3}|^2|1-\re^{-\ri D^2L/2E}|^2$$
 $$=4|U_{i3}|^2|U_{f3}|^2\sin^2(D^2L/4E).$$ Both expressions, for
$f=i$ and $f\neq i$, depend only on two mixing matrix elements and
 not on terms of the form $U_{ik}U_{fl}U_{il}^*U_{fk}^*.$\  Hence,
 there are no CP- or T-violating observations at distances
 $L=o(L_2).$ \ Because of\  $|U_{ij}|^2=|U_{ij}^*|^2$ we have
 $$P_{\overline f \overline i}=P_{fi}\ \ {\rm and}\ \
 P_{if}=P_{fi}\ ,$$
where $P_{\overline f \overline i}$ denotes the probability for
the transition $\overline \nu_i\to\overline\nu_f$.
 The situation changes if $L$ increases and
 approaches  $$L_3\ =\ E/d^2\ .$$
 Here we have
 $$P_{fi}=|U_{i1}U_{f1}^*+U_{i2}U_{f2}^*
\ \re^{-\ri d^2L/2E}+U_{i3}U_{f3}^*\ \re^{-\ri D^2L/2E}|^2$$
 $$=\sum ^3_{j=1}|U_{ij}|^2|U_{fj}|^2+2Re[U_{i1}U_{f2}
U_{i2}^*U_{f1}^*\ \re^{\ri d^2L/2E}$$ $$+
  U_{i1}U_{f3}U_{i3}^*U_{f1}^*\ \re^{\ri D^2L/2E}+U_{i2}U_{f3}
U_{i3}^*U_{f2}^*\ \re^{\ri (D^2-d^2)L/2E}]$$
  which contains three obviously CP-violating contributions if $f\neq i.$ The
  disappearance experiments give CP-symmetric results; for $f=i$ we
  have $$ P_{ii}=\sum
  ^3_{j=1}|U_{ij}|^4+2|U_{i1}|^2|U_{i2}
|^2\cos\frac{d^2L}{2E}+2|U_{i1}|^2|U_{i3}|^2\cos
  \frac{D^2L}{2E}+2|U_{i2}|^2|U_{i3}|^2\cos\frac{(D^2-d^2)L}{2E}$$
$$=\ 1-4[\vert U_{i1}\vert^2\vert U_{i2}\vert^2
   \sin^2\frac{d^2 L}{4E}
+\vert U_{i1}\vert^2\vert U_{i3}\vert^2\sin^2\frac{D^2 L}{4E}
+\vert U_{i2}\vert^2\vert U_{i3}\vert^2\sin^2\frac{(D^2-d^2) L}{4E}]$$
  with the property
  $$P_{\overline i\overline i}=P_{ii}\ .$$\\

  For further discussion of the CP-asymmetries in the appearance
  experiments, it is convenient to introduce factors $p_{ifm}$ 
and phases  $\alpha_{ifm}$ with $m=1,2,3$, defined as
  $$U_{ik}U_{fl}U_{il}^*U_{fk}^*\ =\ \sum_{m=1}^3\ \epsilon_{klm}
 \ p_{ifm}\ \re^{\ri\alpha_{ifm}}\ .$$
The real parts are $p_{ifm}\cos\alpha_{ifm}$, depending on $i$ and $f$,
but independent of the sequence of $ifkl$. The imaginary parts are
$+J^{(\nu)}$ for $ifkl=e\mu 12$ and all cyclic permutations, but
$-J^{(\nu)}$ for all other permutations. 
  In terms of these factors and phases, the appearance rates are
  $$P(i\to f)\ =\ P_{fi}\ =\ \sum^3_{j=1}
|U_{ij}|^2|U_{fj}|^2 \ +\ 2\ \sum_{m=1}^3
\ p_{ifm}\ \cos(\frac{d_m^2 L}{2E}+\alpha_{ifm})$$
$$=\ -4\ \sum _{m=1}^3\ p_{ifm}\ \sin\frac{d_m^2 L}{4E}
\ \sin(\frac{d_m^2 L}{4E}+\alpha_{ifm})\ ,$$
  with  $d_1^2=D^2-d^2$, $d_2^2=D^2$, and $d_3^2=d^2$.
The CP, T, and
CPT symmetry properties of these rates are easily seen; we obtain 
$$P_{\overline f\overline i}\ =\ P_{if}\ =
\ -4\ \sum _{m=1}^3\ p_{ifm}\ \sin\frac{d_m^2 L}{4E}
\ \sin(\frac{d_m^2 L}{4E}-\alpha_{ifm})\ ,$$
which is different
from $P_{fi}$ if at least one of the three phases 
$\alpha_{ifm}$ is different from 0 and
$\pi$. In that case, CP and T are violated, but CPT is conserved.\\

When we approach $L=o(L_3)$,
we have to average over two cosine terms if $D^2\gg d^2$ and if the energy
resolution is limited. We obtain in this case
 $$P_{fi}\ =\ \sum^3_{j=1}|U_{ij}|^2|U_{fj}|^2\ +\ 2\ p_{if3}
 \ \cos(\frac{d^2L}{2E}+\alpha_{if3})\ .$$
In contrast to $L=o(L_2)$, as shown above, there remain CP and T
violating effects at $L=o(L_3)$.\\

It is straightforward to construct T- and CP-violating asymmetries. For
the general case with $L$ between $L_2$ and $L_3$, they are 
$$A_{fi}\ =
\ \frac{P_{fi}-P_{if}}{P_{fi}+P_{if}}\ =\ \frac{P_{fi}-P_{\overline
f\overline i }}{P_{fi}+P_{\overline f\overline i}}\ =\ \frac
{\sum_{m=1}^3\ p_{ifm}\ \sin\alpha_{ifm}
\ \sin(d_m^2 L/4E)\ \cos(d_m^2 L/4E)}
{\sum_{m=1}^3\ p_{ifm}\ \cos\alpha_{ifm}\ \sin^2(d_m^2 L/4E)}\ .$$
For $if=e\mu$, $\mu\tau$, and $\tau e$, the three contributions
$p_{ifm}\sin\alpha_{ifm}$ in the numerator are equal to the 
invariant quantity $J^{(\nu)}$ of the neutrino CKM matrix, leading to
$$ A_{fi}\ =\ J^{(\nu)}\ \frac{\sum_m\ \sin(d^2_mL/4E)
\ \cos(d^2_mL/4E)}
{\sum_m\ p_{ifm}\ \cos\alpha_{ifm}\ \sin^2(d_m^2 L/4E)}\ .$$
For $if=\mu e,\tau\mu,e\tau$, the asymmetries have, obviously,
 the opposite sign. All asymmetries $A_{fi}$ depend on the same
property $J^{(\nu)}$ of the neutrino CKM matrix, a result which can be
found with the same simplicity in the CP violating effects of the
meson sector \cite{JARLSKOG}.\\

Present experiments \cite{SUPERK,SOLAR} have still very large errors on
$\vartheta_{e2}$, $\vartheta_{\mu 3}$, and $\vartheta_{e3}$; but they 
permit to set an upper limit on $J^{(\nu)}$. With
$s_{\mu 3}\approx c_{\mu 3}\approx 1/\sqrt{2}$, $s_{e2}\approx 0.04$, and
$s_{e3}<s_{e2}$, we obtain
$$\vert J^{(\nu)}\vert\ <\ 8\cdot 10^{-4}\ ,$$
whereas the quark sector has the property $J\approx + 3\cdot 10^{-5}$
 \cite{MELE}
if CP violation in $\rm K^0\overline K{}^0$ oscillation has its origin
in a complex quark CKM matrix. The small upper limit on $J^{(\nu)}$
requires a large number of events in future neutrino oscillation
experiments. This paper does not aim to estimate numbers of events. Such
an estimate would require, even without assuming anything on
the phase $\Delta$, better knowledge on the angles
$\vartheta_{e2}$ and $\vartheta_{e3}$.\\

{\Large{\bf Acknowledgement}}\\

 I would like to thank W.\ Hampel (MPI Heidelberg)
und J.\ Urban (TU Dresden) for several helpful discussions.


\begin{thebibliography}{99} 
\itemsep -5pt
\bibitem{SUPERK} Y.\ Fukuda et al.\  (Super-Kamiokande),
        Phys.\ Rev.\ Lett.\ 81 (1998) 1562
\bibitem{SUPERK-2} T.\ Kajita (Super-Kamiokande), presented at
        NEUTRINO 98, Takayama (1998),\\ hep-ex/9810001
\bibitem{CABIBBO} N.\ Cabibbo, Phys.\ Rev.\ Lett.\ 10 (1963) 531
\bibitem{KM} M.\ Kobayashi and T.\ Maskawa, Progr.\ Theor.\ Phys. 
                49 (1973) 652 
\bibitem{PDG} Review of Particle Properties, Particle
           Data Group, Eur.\ Phys.\ J.\ C 3 (1998) 1
\bibitem{JARLSKOG} for a review, see e.\ g.\ C.\ Jarlskog, 
           ``CP Violation'', Advanced Series on Directions in High
           Energy Physics, World Scientific Singapore (1989)
\bibitem{SOLAR} J.\ N.\ Bahcall, P.\ I.\ Krastev, and
           A.\ Yu.\ Smirnov, Phys.\ Rev.\ D 58 (1998) 096016
\bibitem{MSW} L.\ Wolfenstein, Phys.\ Rev.\ D 17 (1978) 2369\\
     S.\ P.\ Mikheyev and A.\ Yu.\ Smirnov, 
       Sov.\ J.\ Nucl.\ Phys.\ 42 (1986) 913
\bibitem{MELE} using $\eta$ from a fit of S.\ Mele, CERN-EP/98-133
         (1998)
    
\end{thebibliography}
\end{document}